\documentclass[journal,twocolumn,10pt]{IEEEtran}
\usepackage{amsmath,amssymb,amsfonts}
\usepackage{graphicx}
\usepackage{cite}
\usepackage{hyperref}
\usepackage{booktabs}
\usepackage{physics}
\usepackage{makecell}
\begin{document}

\title{The Post-Silicon Semiconductor Era: A Review of Physics, Synthesis, and Architectural Integration of Carbon Nanotube Field-Effect Transistors}

\author{
\IEEEauthorblockN{
Suhas Suresh Bharadwaj\textsuperscript{*}, 
Reuben Thomas Thovelil,
Rohith Chembattammal,
Himnish Dave and 
Kabir Jaisinghani
}

\thanks{\textsuperscript{*}Corresponding author; E-mail: f20230029@dubai.bits-pilani.ac.in
}

\vspace{10.5pt}

\IEEEauthorblockA{
Birla Institute of Technology and Science, Pilani -- Dubai Campus
}
}

\maketitle

\begin{abstract}
Silicon CMOS scaling is approaching a set of hard physical limits. Direct source-to-drain quantum tunneling, subthreshold swing degradation, and the Dark Silicon thermal ceiling motivate the search for a new channel material. This review builds the case for single-walled carbon nanotubes (SWCNTs) as that material. We follow a continuous narrative from electronic structure theory through synthesis to integration. The SWCNT bandgap and near-ballistic transport limits are derived from the graphene zone-folding framework and Landauer-B\"{u}ttiker formalism. We benchmark these theoretical limits against ideal coaxial electrostatic bounds to evaluate how the geometry suppresses short-channel effects before tunneling dominates. Comparing this framework against 5 nm experimental data illustrates the aggressive subthreshold degradation driven by source-to-drain tunneling. Furthermore, we derive the exact areal-density equivalence between 1D and 2D quantum capacitance. This demonstrates that close-packed arrays cannot close the dimensional gap to 2D materials, underscoring why superior carrier velocity and electrostatics must carry the CNTFET advantage. Next, we examine CoMoCAT growth and aqueous two-phase extraction against logic fabrication purity demands alongside contact engineering and doping. A closing techno-economic analysis weighs this picture against IEEE IRDS projections. Ultimately, materials purification, contact reliability, and bias temperature instability remain the practical barriers to commercial adoption.
\end{abstract}


\section{Introduction}
\label{sec:introduction}

\IEEEPARstart{S}{ince} Bardeen and Brattain's demonstration of the first point-contact transistor \cite{bardeen1948transistor} and the subsequent transition to silicon as the dominant substrate \cite{silicon1950history}, for over five decades silicon CMOS scaling has been governed by Moore's Law and Dennard scaling \cite{moore1965cramming}. Dennard postulated that scaling physical dimensions, voltage, and current by a constant factor $\kappa > 1$ preserves internal electric fields, maintaining a constant active power density \cite{dennard1974design}:

\begin{equation}
P_{\mathrm{dyn}} = \alpha f C V_{DD}^{2},
\end{equation}

where $\alpha$ is the switching activity factor, $f$ is the clock frequency, $C$ is the total switched capacitance, and $V_{DD}$ is the supply voltage. Under ideal Dennard scaling by $\kappa$, $C \to C/\kappa$, $V_{DD} \to V_{DD}/\kappa$, and $f \to \kappa f$, so $P_{\mathrm{dyn}}/A$ remains constant. However, as modern MOSFETs scale deep into the sub-10~nm regime, these empirical paradigms have collapsed. Because the non-scalable thermal voltage ($k_B T/e$) sets an absolute 300~K minimum subthreshold swing of $60\text{ mV/decade}$, the supply voltage $V_{DD}$ cannot be lowered proportionately without exponentially degrading the $I_{ON}/I_{OFF}$ ratio. Consequently, static leakage now dominates the thermal envelope, inducing the ``Dark Silicon'' phenomenon \cite{esmaeilzadeh2011dark, taur2009fundamentals}. To mitigate this thermal limit, the industry has structurally abandoned planar devices in favor of three-dimensional FinFETs and, more recently, Gate-All-Around (GAA) nanosheets to maximize the gate's electrostatic control over the channel. Despite these major architectural shifts, quantum mechanics is the primary reason silicon breaks down at these scales.

\begin{figure}[!t]
    \centering
    \includegraphics[width=1\columnwidth]{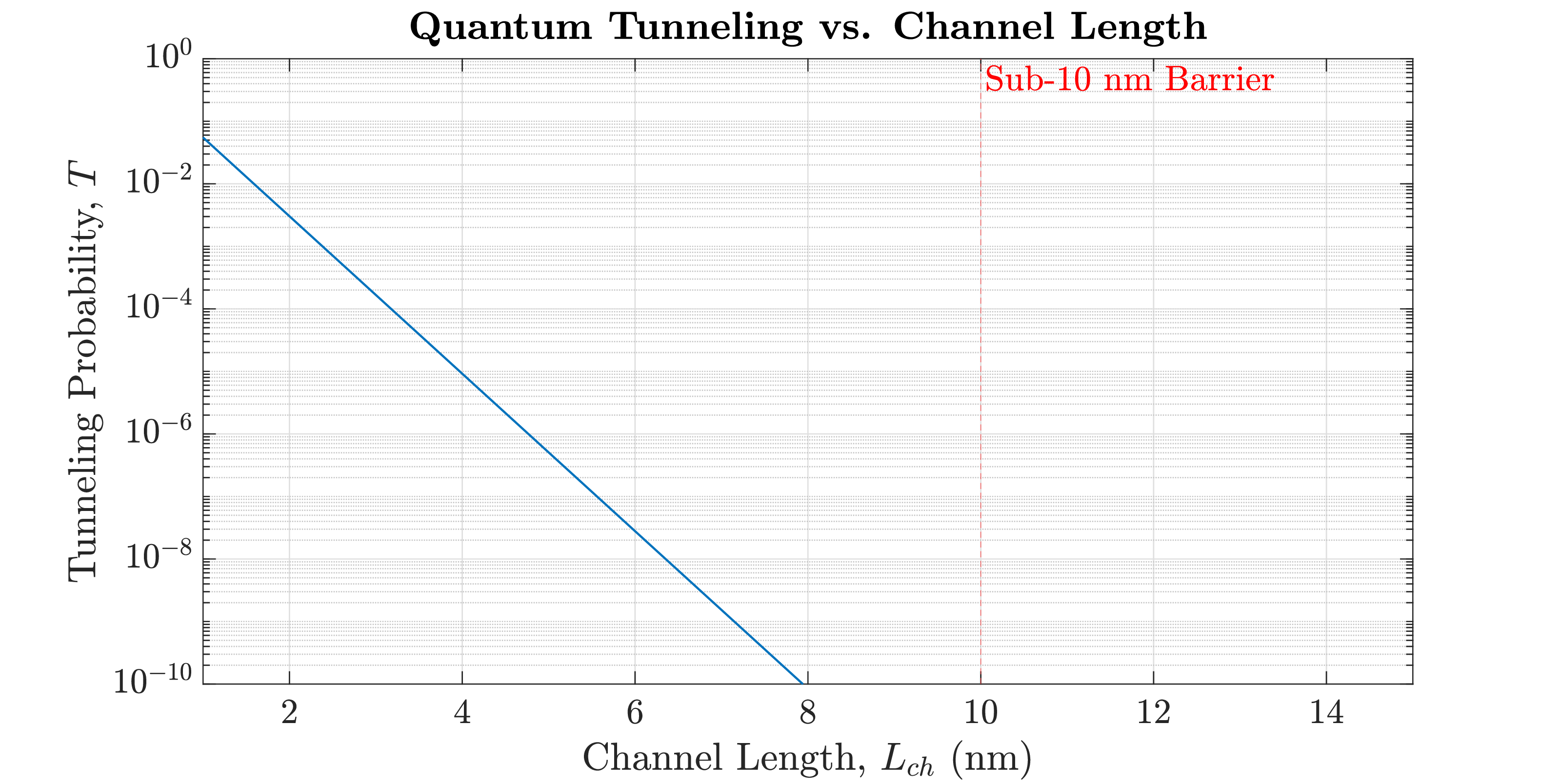}
    \caption{Phenomenological WKB tunneling probability $T$ vs.\ channel length $L_{ch}$ for direct source-to-drain tunneling. This serves purely as a generic silicon pedagogical illustration using a parabolic channel barrier, parameterized with the lighter Si transverse mass ($m^{*}=0.19\,m_0$ where $m_0$ is the bare electron mass) and a representative barrier height ($U-E=0.5$~eV). Applying this directly to SWCNTs requires replacing the parabolic assumption with a linear dispersion model and full 3D Density of States integration across the transport direction. Therefore, this serves as a qualitative first-order illustration of the exponential tunneling surge rather than a quantitative nanotube extraction~\cite{taur2009fundamentals,sze2006physics}.}
    \label{fig:tunneling}
\end{figure}

When the channel length ($L_{ch}$) or the physical thickness of the gate dielectric approaches the de Broglie wavelength of a conduction electron, classical electrostatic potential barriers no longer act as absolute, rigid walls. While gate leakage via Fowler-Nordheim tunneling was historically mitigated by high-$\kappa$ dielectrics (e.g., $\text{HfO}_2$) replacing $\text{SiO}_2$~\cite{taur2009fundamentals}, as $L_{ch}$ shrinks below 5~nm electrons begin to directly tunnel from source to drain through the off-state potential barrier. Using the Wentzel-Kramers-Brillouin (WKB) approximation, this direct transmission probability $T$ can be estimated as:

\begin{equation}
T \approx \exp\left(-\frac{2}{\hbar} \int_{0}^{L_{ch}} \sqrt{2m^* (U(x) - E)} \, dx\right),
\label{eq:wkb}
\end{equation}

where $\hbar$ is the reduced Planck constant, $m^*$ is the carrier's effective mass in the lattice, $U(x)$ is the barrier potential, and $E$ is the incident electron energy \cite{taur2009fundamentals}. As illustrated in Fig.~\ref{fig:tunneling}, because the channel length $L_{ch}$ directly dictates the integration bounds in the exponent, any further physical reduction in $L_{ch}$ causes an exponential surge in quantum tunneling. This generates massive off-state leakage currents ($I_{off}$), fundamentally destroying the transistor's ability to act as an effective binary switch.

Simultaneously, devices at this scale suffer from severe short-channel effects (SCEs). The subthreshold swing ($SS$) itself is governed by the capacitive divider between the gate and the channel body:
\begin{equation}
SS = \ln(10) \frac{k_B T}{e} \left( 1 + \frac{C_{dep}}{C_{ox}} \right),
\end{equation}
where $C_{dep}$ is the depletion (body) capacitance that competes with the gate for control of the channel potential \cite{sze2006physics}. A poorly controlled gate lets $C_{dep}$ dominate, pushing $SS$ above the $300$~K thermionic floor of $\ln(10)\,k_BT/e \approx {60}~\text{mV/decade}$.

A related effect, drain-induced barrier lowering (DIBL), arises when the drain field penetrates the channel via overlapping depletion regions, lowering the barrier and reducing $V_T$ roughly linearly with $V_{DS}$ \cite{troutman1979vlsi,taur2009fundamentals}. The conformal, all-sides gate field of a Gate-All-Around geometry minimizes both $C_{dep}/C_{ox}$ and this drain-channel coupling, making GAA the natural electrostatic target for CNTFET integration.

Circumventing these limits without thermal runaway requires ultra-thin, low-dimensional channel materials that maximize gate capacitance while minimizing the cross-section available for off-state leakage. Single-walled carbon nanotubes (SWCNTs), formed by rolling a single-atom-thick graphene sheet into a seamless cylinder, approach this limit by confining electrons radially and forcing transport into a strictly one-dimensional (1D) regime \cite{iijima1993single, bethune1993cobalt, avouris2002carbon}.

\begin{table*}[!t]
\centering
\caption{Quantitative comparison of post-silicon channel materials. Ranges reflect the spread of reported values in the cited literature. $v_{inj}$ denotes the quasi-ballistic injection velocity at the source barrier, distinct from the phonon-limited saturation velocity $v_{sat}$ used for SWCNTs in Section~\ref{sec:phonon-emission}. SWCNT and InAs NW share comparable $v_{inj}$, but the SWCNT advantage derives from superior gate electrostatics, BEOL compatibility, and the absence of lattice-mismatch buffers. The SWCNT SS range spans from the 5~nm result of Qiu~\emph{et~al.}~\cite{qiu2017scaling} to older geometries.}
\label{tab:materials}
\begin{tabular}{@{}lccccl@{}}
\toprule
\textbf{Material} & \textbf{Peak $v_{inj}$ (cm/s)} &
\textbf{Best Exp.\ SS (mV/dec)} & \textbf{BEOL ($<$400\textdegree C)} &
\textbf{Gate Electrostatics} & \textbf{Refs.} \\ \midrule
InAs NW          & ${\sim}4\times10^7$ & ${\sim}65$ (GAA) & No  & \makecell[l]{Good in gate-all-around geometry, \\ poor in planar/back-gated geometry} & \cite{chuang2013ballistic} \\
Si (FinFET)      & ${\sim}1\times10^7$ & 65--70 (Tri-gate) & No  & Good & \cite{auth201222nm} \\
MoS$_2$ (mono.)  & ${\sim}3\times10^6$ & ${\sim}65$ (Top-gate) & Yes & Good & \cite{desai2016mos2} \\
\textbf{1D SWCNT} & ${\sim}4\times10^7$ & 73--94 & Yes & Excellent & \cite{franklin2012sub10,qiu2017scaling} \\ \bottomrule
\end{tabular}
\end{table*}

This 1D confinement has profound transport implications. In bulk silicon (3D), electrons scatter in all directions when interacting with acoustic lattice vibrations (phonons). In a 1D SWCNT, the phase space for scattering is severely restricted as an electron can only scatter forward or backward. Since most acoustic phonons lack the necessary momentum to completely reverse an electron's trajectory, backscattering is drastically suppressed. Consequently, the mean free path ($l_m$) of carriers in a pristine SWCNT easily exceeds $1~\mu m$. Since modern channel lengths are strictly $L_{ch} \ll l_m$, carrier transport approaches the absolute ballistic limit. Following the Landauer-B\"{u}ttiker formalism for quantum transport, the conductance $G$ of a 1D ballistic channel is quantized:

\begin{equation}
G = \frac{2e^2}{h} M \cdot T.
\label{eq:landauer}
\end{equation}

Because semiconducting SWCNTs exhibit two valley-degenerate conducting channels ($M=2$, corresponding to the $K$ and $K'$ valleys of graphene, with spin degeneracy already incorporated in the $2e^2/h$ prefactor) with a transmission probability $T \approx 1$, their theoretical quantum conductance limit is $4e^2/h \approx 155~\mu\text{S}$, allowing them to sustain large on-currents at low operating voltages \cite{datta1997electronic}. Quantized conductance steps consistent with the general Landauer picture (though measured in multi-walled, not single-walled, nanotubes) have been directly observed experimentally \cite{Frank1998}.

Furthermore, the physical geometry of an SWCNT (diameter $d_{t} \approx 1$--$2~nm$) drastically improves the transistor's natural scaling length ($\lambda$), which governs the distance over which the gate's electric field decays in the channel:

\begin{equation}
\lambda_{\mathrm{GAA}} = \frac{d_t}{2}
  \sqrt{\frac{\varepsilon_{\mathrm{ch}}}{\varepsilon_{\mathrm{ox}}}
  \ln\!\left(1 + \frac{2t_{\mathrm{ox}}}{d_t}\right)},
\label{eq:lambda_GAA}
\end{equation}

where $d_t$ is the tube diameter, $\varepsilon_{\mathrm{ch}}$ is the effective dielectric permittivity of the SWCNT channel interior (approximated as $\varepsilon_{\mathrm{ch}} \approx \varepsilon_0$ for the hollow nanotube core in the coaxial Poisson model, neglecting the tube shell's own electronic polarizability~\cite{Marchi2006,rahman2003theory}), $\varepsilon_{\mathrm{ox}}$ is that of the gate dielectric, and $t_{\mathrm{ox}}$ is the dielectric thickness. This expression is derived from the cylindrical Poisson equation for a coaxial-gate geometry~\cite{Marchi2006, rahman2003theory} and differs substantially from the planar approximation $\lambda_{\mathrm{planar}} = \sqrt{\varepsilon_{\mathrm{ch}} t_{\mathrm{ch}} t_{\mathrm{ox}} / \varepsilon_{\mathrm{ox}}}$ at the nanometer-scale dimensions relevant here.

Table~\ref{tab:materials} positions SWCNTs among the broader landscape of beyond-silicon channel materials. 2D transition metal dichalcogenides (e.g., MoS$_2$) offer comparable atomic body thickness and even sub-1~nm gate-length demonstrations \cite{radisavljevic2011single,chhowalla2016two,desai2016mos2}, but suffer from lower carrier mobility and poor Ohmic contacts \cite{allain2015electrical}. III--V nanowires (e.g., InAs) provide excellent injection velocities \cite{del2011nanometre} but require lattice-mismatch buffer layers that preclude direct monolithic Si stacking \cite{ko2010ultrathin} and exhibit poor planar electrostatics. Stacked Si GAA nanosheets \cite{loubet2017stacked} maximize electrostatic containment but cannot resolve 3D acoustic phonon scattering. Of all alternatives, SWCNTs uniquely combine atomic body thickness, near-ballistic transport, and low-temperature deposition within one material system \cite{franklin2015nanomaterials}.

Choosing SWCNTs over these alternatives, however, does not by itself guarantee a working transistor: the same zone-folding condition that fixes the bandgap in Section~\ref{sec:physics} also fixes the tube's electrical character, and it does so indiscriminately. The $(n-m) \equiv 0\pmod{3}$ condition statistically applies to approximately one-third of all possible $(n,m)$ chiral indices, making the metallic tube fraction approximately 33\% in statistically uncontrolled growth \cite{saito1998physical}. This is the fundamental motivation for the purification protocols detailed in Section~\ref{sec:manufacturing}. Prior CNTFET reviews~\cite{avouris2003carbon,zhangpeng2019digital} predate the IRDS 2022 roadmap, wafer-scale ATPE purity data, monolithic 3D BEOL integration~\cite{yuan2023three}, and all published bias-temperature-instability characterization~\cite{sun2025bti}. This review closes that gap by synthesizing these post-2019 results into a unified physical and techno-economic framework. Because the aim of this review is to construct a continuous narrative from band theory to foundry economics, several intermediate chemical and manufacturing concepts are represented via standardized qualitative schematics. This allows the quantitative focus to remain squarely on the electrostatic and quantum scaling limits. While the title introduces a broad post-silicon framework, the specific focus of this text remains rigorously on carbon nanotubes.

Sections~\ref{sec:physics}--\ref{sec:conclusion} proceed from SWCNT electronic structure, through ballistic transport and quantum-capacitance limits, to synthesis, device integration, techno-economic feasibility, and open challenges.

\section{Fundamental Electronic Structure of SWCNTs}
\label{sec:physics}
 
Rather than re-deriving the graphene tight-binding model (see \cite{saito1998physical}), this section states the electronic-structure results that govern CNTFET performance.
 
\subsection{Chirality, Diameter, and the Metallic/Semiconducting Dichotomy}
 
A SWCNT is fully characterized by its chiral index pair $(n,m)$ \cite{saito1998physical}, which defines the circumferential vector $\mathbf{C}_h = n\mathbf{a}_1 + m\mathbf{a}_2$ on the parent graphene lattice and fixes the tube diameter as $d_t = a\sqrt{n^2 + nm + m^2}/\pi$ (where lattice constant $a \approx 2.46$~\AA). Imposing periodic boundary conditions along $\mathbf{C}_h$ discretizes the allowed transverse wave-vectors into parallel 1D lines within the 2D graphene Brillouin zone (zone-folding approximation \cite{saito1998physical}). The electrical character of the tube is governed entirely by whether any allowed $k$-line intersects the $K/K'$ Dirac points, yielding metallic behavior when $n - m = 3q$ (where $q$ is an integer) and semiconducting behavior otherwise. For semiconducting tubes, linearizing the graphene dispersion near the Dirac point yields the fundamental bandgap \cite{ando2005theory}:

\begin{equation}
  E_g = \frac{2\gamma_0 a_{cc}}{d_t},
  \label{eq:bandgap}
\end{equation}

where $\gamma_0 \approx 2.7$~eV is the nearest-neighbor tight-binding overlap integral and $a_{cc} \approx 1.42$~\AA\ is the carbon--carbon bond length. For the representative $(13,0)$ zigzag tube ($d_t \approx 1.018$~nm), this yields $E_g \approx 0.75$~eV. As an internal consistency check, subtracting the electron affinity ($\chi \approx 4.3$~eV) from the ionization potential ($IP \approx 5.05$~eV) independently reported by Jiang~\emph{et~al.}~\cite{jiang2007quasiparticle} recovers this exact 0.75~eV bandgap.

This relationship imposes a specific constraint. Based on the diameter-dependent quasiparticle corrections to the band edges reported by Jiang~\emph{et~al.}~\cite{jiang2007quasiparticle}, we determine that below $d_t<0.8$~nm, the ionization potential is pushed too high for Pd ($\Phi_M\approx5.10$~eV) to yield a near-zero hole barrier. Conversely, above $d_t>1.4$~nm, $E_g<0.55$~eV worsens ambipolar leakage. The optimal logic-grade range is therefore $d_t=0.9$--$1.3$~nm.

Because the dominant CoMoCAT chiralities (e.g., $(6,5)$ at $0.75$~nm) fall below this optimal window, their integration necessitates accepting higher Schottky barriers or post-synthesis diameter sorting. A $(13,0)$ zigzag tube ($d_t \approx 1.018$~nm, $E_g \approx 0.75$~eV) therefore serves as a consistent analytical baseline for the theoretical limits discussed in Sections~\ref{sec:transport} and~\ref{sec:integration}. It provides a well-characterized bandgap devoid of the curvature-induced $\sigma$--$\pi$ rehybridization most pronounced in near-metallic tubes, which scales as $E_g^{\mathrm{mini}} \propto 1/d_t^2$ and remains thermally negligible ($<5$~meV) for $d_t>2$~nm but can reach tens-to-hundreds of meV for $d_t<1$~nm \cite{kane1997size,ando2005theory}.
 
\subsection{1D Density of States: Van Hove Singularities}
 
\begin{figure}[!t]
    \centering
    \includegraphics[width=1\columnwidth]{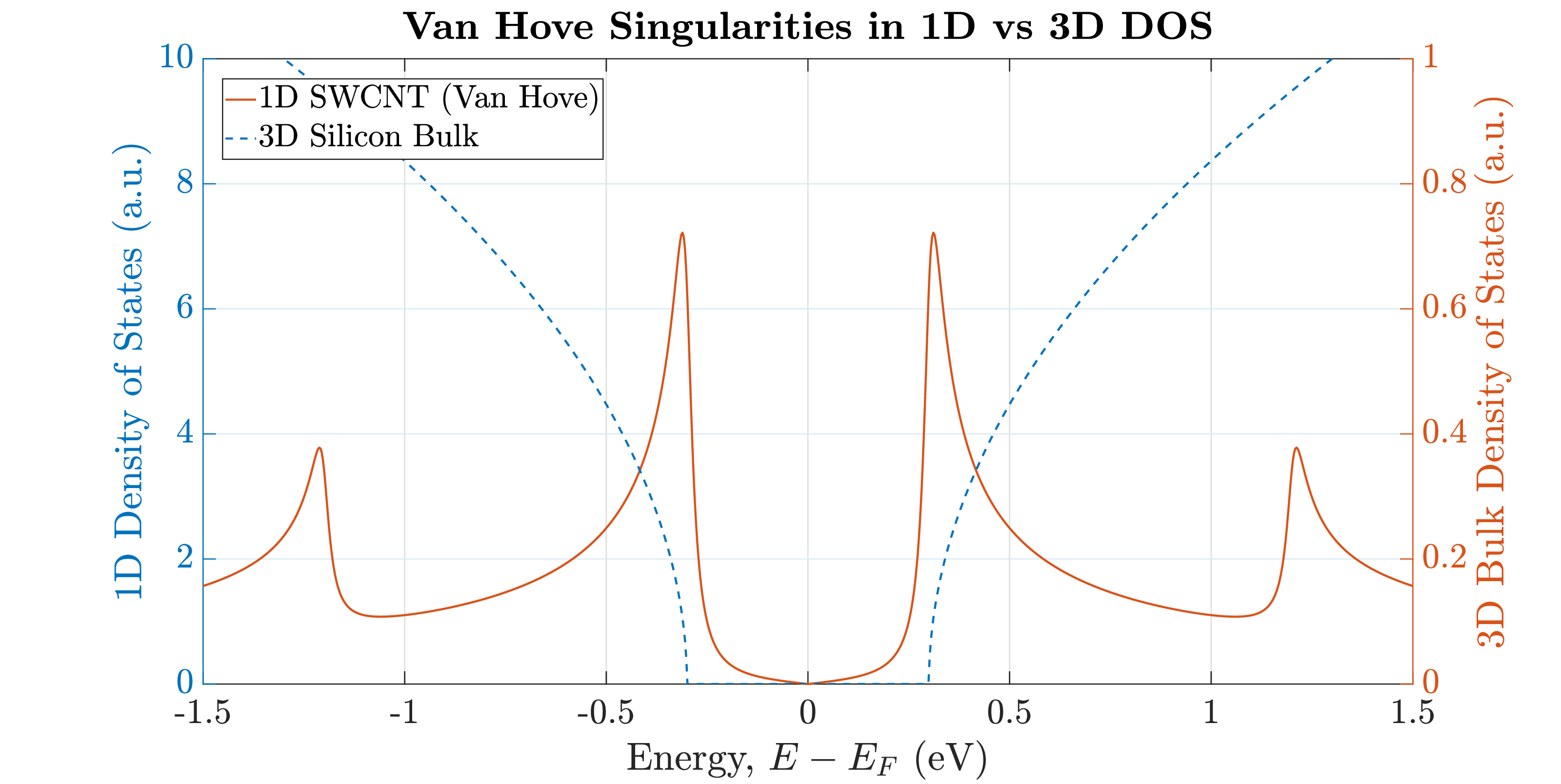}
    \caption{Analytical 1D DOS from Eq.~\eqref{eq:dos} for a $(13,0)$ SWCNT ($E_g\approx0.75$~eV, $\gamma_0=2.7$~eV). A qualitative, arbitrary-unit representation of a 3D bulk-Si square-root DOS is overlaid on a dual scale purely for schematic comparison. This highlights the divergent Van Hove singularities unique to 1D confinement against a traditional continuous background.}
    \label{fig:dos}
\end{figure}

The strict 1D confinement produces a Density of States that is fundamentally different from the smooth, continuous square-root DOS of 3D bulk silicon. The 1D DOS per unit length is:

\begin{equation}
  D(E) = \frac{4}{\pi \hbar v_F}
  \sum_{i} \frac{|E|}{\sqrt{E^2 - E_i^2}}\,\Theta(|E|-E_i),
  \label{eq:dos}
\end{equation}

where $E$ is measured from the charge-neutrality (Dirac) point, $E_i$ denotes the energy of the $i$-th subband onset measured from the same origin, $\Theta$ is the Heaviside step function, and the prefactor of 4 accounts for the two-fold spin and two-fold valley ($K/K'$) degeneracy~\cite{mintmire1998universal,saito1998physical}. As $E \to E_i$, the denominator diverges, producing the sharp Van Hove singularities (VHSs) shown in Fig.~\ref{fig:dos}. When the gate voltage shifts $E_F$ into a VHS, a massive carrier density becomes available instantaneously, enabling the exceptionally high $I_{\mathrm{ON}}$ that distinguishes CNTFETs from planar silicon devices.
 
\subsection{Phonon Phase Space Restriction and Near-Ballistic Transport}
\label{sec:phonon}
 
The most technologically important consequence of 1D confinement is not the Van Hove DOS itself, but its severe restriction of the electron-phonon scattering phase space. In a 3D silicon lattice, an electron can scatter into any point on a spherical Fermi surface via low-energy acoustic phonons. In a 1D SWCNT, the scattering phase space collapses to a single axis, and backscattering now requires a phonon carrying momentum $q = 2k_F$. Intravalley ($K \to K$) acoustic backscattering is further suppressed by parity selection rules, and intervalley ($K \to K'$) scattering requires zone-boundary optical phonons that are statistically depopulated at low biases \cite{suzuura2002phonons, ando2005theory}. The low-field mean free path consequently exceeds $1~\mu$m in pristine SWCNTs \cite{white1998carbon}, rendering transport in sub-10~nm logic channels near-ballistic. As quantified by Eq.~\eqref{eq:lambda_GAA}, the atomic-scale tube diameter ($d_t \approx {1}~{nm}$) minimizes $\lambda_{\mathrm{GAA}}$, granting the coaxial GAA gate near-perfect electrostatic suppression of DIBL at channel lengths well below {10}~{nm}~\cite{franklin2012sub10}.

\section{Quantum Transport Physics and Ballistic Scaling Limits}
\label{sec:transport}

While the static bandgap dictates subthreshold switching, dynamic $I_{ON}$ is governed by non-equilibrium quantum transport. The near-ballistic transport established qualitatively in Section~\ref{sec:phonon} follows from Fermi's Golden Rule, $\Gamma_{\mathbf{k}\to\mathbf{k}'}=(2\pi/\hbar)|\langle\mathbf{k}'|H_{ep}|\mathbf{k}\rangle|^2\delta(E_{\mathbf{k}'}-E_{\mathbf{k}}-s\hbar\omega_q)$, where $s=\pm1$ denotes absorption/emission~\cite{ando2005theory,suzuura2002phonons}: since the $K$ and $K'$ valleys are separated by $\Delta k=2|K|$, intervalley scattering requires a momentum transfer far exceeding that available from low-energy phonons, reinforcing the near-elimination of backscattering \cite{suzuura2002phonons}.

\subsection{High-Field Transport and Optical Phonon Emission}
\label{sec:phonon-emission}

The near-elimination of backscattering established above holds only in the low-field limit. Under high drain biases ($V_{DS} \sim 0.7 \text{ V}$), electrons instead gain sufficient kinetic energy to overcome the optical phonon emission threshold ($\hbar\omega_{OP} \approx 160 \text{ meV}$)~\cite{perebeinos2005electron}. This triggers highly efficient intervalley scattering, capping the drift velocity at $v_{sat} \approx 3$--$4\times10^7$~cm/s. This optical-phonon emission mechanism imposes an absolute saturation current limit of $I_{\mathrm{max}} \approx G_Q \cdot \hbar\omega_{OP}/e \approx 25$--$30~\mu\text{A}$ per tube~\cite{yao2000high,perebeinos2005electron}. Realistic per-tube on-currents, which can reach $\sim$15--20~$\mu$A with optimized contacts (Section~\ref{sec:integration}), remain comfortably below this theoretical ceiling. Reaching mA/$\mu$m drive currents competitive with silicon FinFETs therefore requires parallel arrays of 50--200 tubes/$\mu$m, a density target that motivates the assembly strategies of Section~\ref{sec:manufacturing}.

Three carrier velocities matter here: the intrinsic Fermi velocity $v_F \approx 8\times10^7$~cm/s (band structure, quantum capacitance), the quasi-ballistic injection velocity $v_{inj} \approx 4\times10^7$~cm/s (thermally broadened source injection), and the phonon-limited saturation velocity $v_{sat}$ (maximum drift current under high bias).

\subsection{The Intrinsic Capacitance Limit}

Section~\ref{sec:phonon-emission} bounded the CNTFET's maximum drive current. Its switching speed and energy are bounded by a second independent constraint. In conventional MOSFETs, the total gate capacitance is assumed equivalent to the oxide capacitance ($C_{ox}$). However, because SWCNTs possess a highly restricted 1D Density of States (DOS), the channel itself possesses a finite ability to absorb charge, known as the quantum capacitance ($C_Q$).

The channel's finite ability to absorb charge is dictated by its intrinsic capacitance per unit length. As originally derived by Burke using a Luttinger liquid and transmission-line framework~\cite{burke2002luttinger}, the intrinsic gate capacitance is directly proportional to the localized density of states. For the linear SWCNT dispersion with four-fold degeneracy, this yields an intrinsic capacitance limit:

\begin{equation} 
\frac{C_Q}{L} = \frac{4e^2}{\pi\hbar v_F} \approx 390~\text{aF}/\mu\text{m},  
\label{eq:cq_per_length} 
\end{equation}  

evaluated at the Fermi velocity $v_F = 8\times10^7$~cm/s. It must be noted that under high drain bias ($V_{DS} \approx V_{DD}$), the drain-side density of states is pulled down, rendering $C_Q$ spatially non-uniform across the channel.

Since $C_{tot}^{-1}=C_{ox}^{-1}+C_Q^{-1}$, $C_{tot}$ bottlenecks at $C_Q$ as foundries drive $C_{ox}\to\infty$ with ultra-thin high-$\kappa$ dielectrics, yielding diminishing returns on $g_m$. Dielectric scaling must therefore be optimized against the $C_Q$ limit of the specific $(n,m)$ lattice to fully exploit ballistic transport.

$C_Q$ is also strongly gate-voltage dependent. It is near-zero when $E_F$ sits in the bandgap, in the off state, and rises steeply as $E_F$ approaches the first Van Hove singularity at threshold, which is exactly where the series bottleneck is most severe and where CNTFET logic spends most of its operating time. Dielectric scaling therefore yields its largest switching-energy benefit at low gate overdrive, consistent with the low-$V_{DD}$ operation envisioned for CNTFET standard cells \cite{ilani2006measurement}.

This quantum capacitance framework also establishes the absolute kinematic speed limit of the device. While Section~\ref{sec:phonon-emission} bounds the continuous drive current via optical phonon emission, the dynamic RC switching delay is bounded by an independent quantum-mechanical constraint dictated by the Landauer conductance quantum $G_{Q} = 4e^{2}/h$~\cite{datta1997electronic}. Evaluating the intrinsic channel delay ($\tau_{int} = (C_{Q}/L)\,L_{ch}/G_{Q}$) using Eq.~\eqref{eq:cq_per_length}, with $G_Q = 4e^2/h = 2e^2/(\pi\hbar)$, the fundamental constants partially cancel:

\begin{equation}
\tau_{int} = \frac{(C_Q/L)\, L_{ch}}{G_Q} = \frac{\dfrac{4e^2 L_{ch}}{\pi\hbar v_F}}{\dfrac{2e^2}{\pi\hbar}} = \frac{2L_{ch}}{v_F},
\label{eq:transit_time}
\end{equation}

where the factor of two arises because $C_Q/L$ counts both forward- and backward-moving electronic states in the one-dimensional channel---a phenomenon rigorously treated in Burke's distributed kinetic-inductance transmission-line model~\cite{burke2002luttinger}---while the unidirectional ballistic transit time is $\tau_{\mathrm{transit}} = L_{ch}/v_F$~\cite{datta1997electronic}. The intrinsic RC delay $\tau_{int}$ therefore equals twice the ballistic transit time, an inherent consequence of the channel charging both directions simultaneously. This establishes the maximum achievable intrinsic transit frequency as $f_{T,\mathrm{max}} = 1/(2\pi\tau_{int}) = v_F/(4\pi L_{ch})$, which for $L_{ch} = 10$~nm and $v_F = 8\times10^7$~cm/s gives $f_{T,\mathrm{max}} \approx 6.4$~THz, the absolute kinematic ceiling for a device at this channel length, consistent with Burke's independent projection of ballistic THz-regime operation in scaled nanotube transistors~\cite{burke2004ac}. By normalizing the Landauer conductance against the 1D density of states, it becomes analytically evident that this fundamental RC switching delay is entirely decoupled from physical dielectric scaling.

\subsection{Dimensionality and the Array Equivalence Limit}

Most contemporary literature reviews group 1D nanotubes and 2D transition metal dichalcogenides together as generic ultra-thin body candidates. This grouping overlooks a fundamental mathematical divergence in their electrostatics. By comparing the quantum capacitance of a 1D SWCNT with a continuous 2D monolayer, we can derive the exact density required for a nanotube array to match the charge-carrying capacity of a 2D sheet.

For a 2D semiconductor like MoS$_2$ exhibiting a parabolic band structure, the density of states is independent of energy. This yields a constant 2D quantum capacitance per unit area defined by the effective mass $m^*$ and the valley degeneracy $g_v$:

\begin{equation}
  C_{Q,\text{2D}} = \frac{e^2 g_v m^*}{\pi \hbar^2}.
  \label{eq:cq_2d}
\end{equation}

Assuming $g_v=2$ and an effective mass of $m^* \approx 0.5 m_0$ for MoS$_2$, this produces an intrinsic area capacitance of approximately $66.9~\mu\text{F}/\text{cm}^2$~\cite{chhowalla2016two}.

In contrast, the 1D quantum capacitance of a single SWCNT is fixed by the Fermi velocity as established previously in Eq.~\eqref{eq:cq_per_length}. To determine the critical packing density $\rho_{\text{crit}}$ where an array of 1D tubes matches the capacitive strength of a continuous 2D sheet, we equate the area-normalized 1D capacitance to the 2D limit:

\begin{equation}
  \rho_{\text{crit}} \cdot C_{Q,\text{1D}} = C_{Q,\text{2D}}.
\end{equation}

Because MoS$_2$ has a valley degeneracy of two and a spin degeneracy of two, and the SWCNT possesses parallel two-fold degeneracies, this substitution yields a straightforward analytical equivalence point:

\begin{equation}
  \rho_{\text{crit}} = \frac{m^* v_F}{2 \hbar}.
  \label{eq:rho_crit}
\end{equation}

Evaluating this expression with an effective mass $m^* \approx 0.5 m_0$ yields a required density of approximately $1720~\text{tubes}/\mu\text{m}$. Because the physical diameter of a single tube is roughly $1~\text{nm}$, the maximum theoretical close-packing limit is $1000~\text{tubes}/\mu\text{m}$. This pedagogical synthesis demonstrates that an SWCNT array cannot physically equal the raw charge-carrying footprint of a 2D monolayer. Instead, it provides a clear mathematical justification for why high-density assembly targets are mandatory. CNTFETs must offset this geometric capacitance penalty by leveraging their higher injection velocity and conformal GAA electrostatics.

\subsection{Electrostatic Scaling Limits and Dynamic Power Efficiency}
 
To translate the equations of Sections~\ref{sec:physics}--\ref{sec:transport} into a predictive logic baseline, we establish an idealized reference geometry at the 10~nm node. This node is chosen as a threshold inferred from the sub-5~nm tunneling onset reported empirically below, rather than derived from a CNT-specific WKB calculation. Dielectric efficiency is measured by the Equivalent Oxide Thickness:

\begin{equation}
\mathrm{EOT} = t_{\mathrm{hk}} \left( \frac{\varepsilon_{\mathrm{SiO_2}}}{\varepsilon_{\mathrm{hk}}} \right).
\label{eq:EOT_def}
\end{equation}

For a standard 2~nm conformal HfO$_2$ layer ($\varepsilon_r \approx 25$), this gives $\mathrm{EOT} \approx 0.31$~nm, well within the sub-1~nm EOT target of the IRDS Beyond CMOS roadmap \cite{irds2022}. Coupled with the optimal natural scaling length ($\lambda_{\mathrm{GAA}}$) of the 1D atomic cylinder, this allows the subthreshold swing to theoretically converge to the 300~K thermionic limit ($\ln(10)\,k_BT/e \approx 59.6~\text{mV/decade}$) in the absence of traps, a limit corroborated by advanced Non-Equilibrium Green's Function (NEGF) treatments of ideal GAA geometries~\cite{rahman2003theory}.

It is critical to note that while Eq.~\eqref{eq:lambda_GAA} establishes the theoretical ideal for a wrap-around coaxial gate, all experimentally cited devices in this review utilize planar or top-gated geometries. Consequently, the subthreshold degradation observed in sub-10~nm experimental devices stems from a combination of direct source-to-drain tunneling and non-ideal planar fringing fields, rather than tunneling alone.

However, as $L_{ch}$ is physically scaled down to 5~nm in state-of-the-art fabricated devices, the WKB tunneling probability (Eq.~\eqref{eq:wkb}) begins to manifest. As demonstrated experimentally by Qiu~\emph{et~al.} at a 5~nm gate length, the subthreshold swing degrades to $S \approx 73~\text{mV/decade}$ even with optimized ultrathin graphene contacts \cite{qiu2017scaling}. Interestingly, Qiu~\emph{et~al.} achieved a superior subthreshold swing at 5~nm compared to earlier 10~nm devices due to their utilization of ultrathin graphene source and drain electrodes, which provided exceptionally transparent carrier injection. Bridging the analytical 10~nm baseline with this 5~nm empirical data strongly supports the hypothesis that direct source-to-drain tunneling, rather than classical short-channel electrostatics, remains the ultimate physical boundary for 1D scaling.

Beyond subthreshold control, 1D confinement profoundly impacts dynamic switching energy, proportional to the Power-Delay Product (PDP $\propto C_{tot} V_{DD}^2$). While highly quantum-confined silicon devices, such as the 1~nm GAA silicon nanowire analyzed by Wang~\emph{et~al.}~\cite{wang2004quantum}, are also heavily bottlenecked by quantum capacitance, the 1D SWCNT maintains a stricter fundamental limit ($C_Q \approx 390~\text{aF}/\mu\text{m}$). This intrinsic capacitance ceiling suggests that CNTFETs should consume less dynamic switching energy than bulk 3D devices at similar pitches. However, interconnect and fan-out parasitics will ultimately dictate standard-cell performance.

\subsection{Intrinsic Transit Frequency and RF Figure of Merit}
 
The intrinsic unity-current-gain frequency $f_T = g_m/(2\pi C_{tot})$ characterizes the CNTFET's fundamental speed limit independently of circuit parasitics. In the ballistic limit, where transconductance is bounded by the conductance quantum ($g_m \approx G_Q$) and capacitance is bottlenecked by the density of states ($C_{tot} \approx C_Q$), this standard AC small-signal relation algebraically collapses into the exact kinematic transit-time ceiling derived in Eq.~\eqref{eq:transit_time}:

\begin{equation}
  f_T \approx \frac{G_Q}{2\pi (C_Q/L) L_{ch}} = \frac{v_F}{4\pi L_{ch}}.
  \label{eq:ft}
\end{equation}

At a 10~nm channel length, this evaluates identically to the 6.4~THz limit. This demonstrates that beyond a certain point, physical dielectric scaling ($C_{ox} \to \infty$) can no longer accelerate a ballistic CNTFET.

Parasitic capacitances from gate overlap, pad fringing, and routing reduce this to an \emph{extrinsic} $f_T$ that is one to two orders of magnitude lower, in the tens-to-hundreds-of-GHz range \cite{rutherglen2009nanotube}. This is still competitive with advanced FinFETs, and it has been experimentally confirmed above 100~GHz for wafer-scalable, self-aligned CNTFETs~\cite{rutherglen2019wafer}, with self-aligned-gate architectures subsequently pushing measured $f_T$ to 240~GHz~\cite{xie2023carbon}, and a recent preprint reports electro-thermal co-design enabling flexible RF CNTFETs exceeding 100~GHz, pending independent peer review~\cite{xia2025flexible}. GHz ring oscillators and frequency doublers on wafer-scale CNT arrays, with toggle frequencies exceeding 5~GHz~\cite{zhong2018gigahertz}, confirm that this advantage translates to circuit-level switching, provided purified, aligned arrays of sufficient density are available. That requirement is the subject of the next section.

\section{Wafer-Scale Synthesis, Deterministic Assembly, and High-Density Integration}
\label{sec:manufacturing}

\begin{figure*}[!t]
    \centering
    \begin{minipage}{\dimexpr0.5\textwidth-0.1in\relax}
        \centering
        \includegraphics[width=1\linewidth]{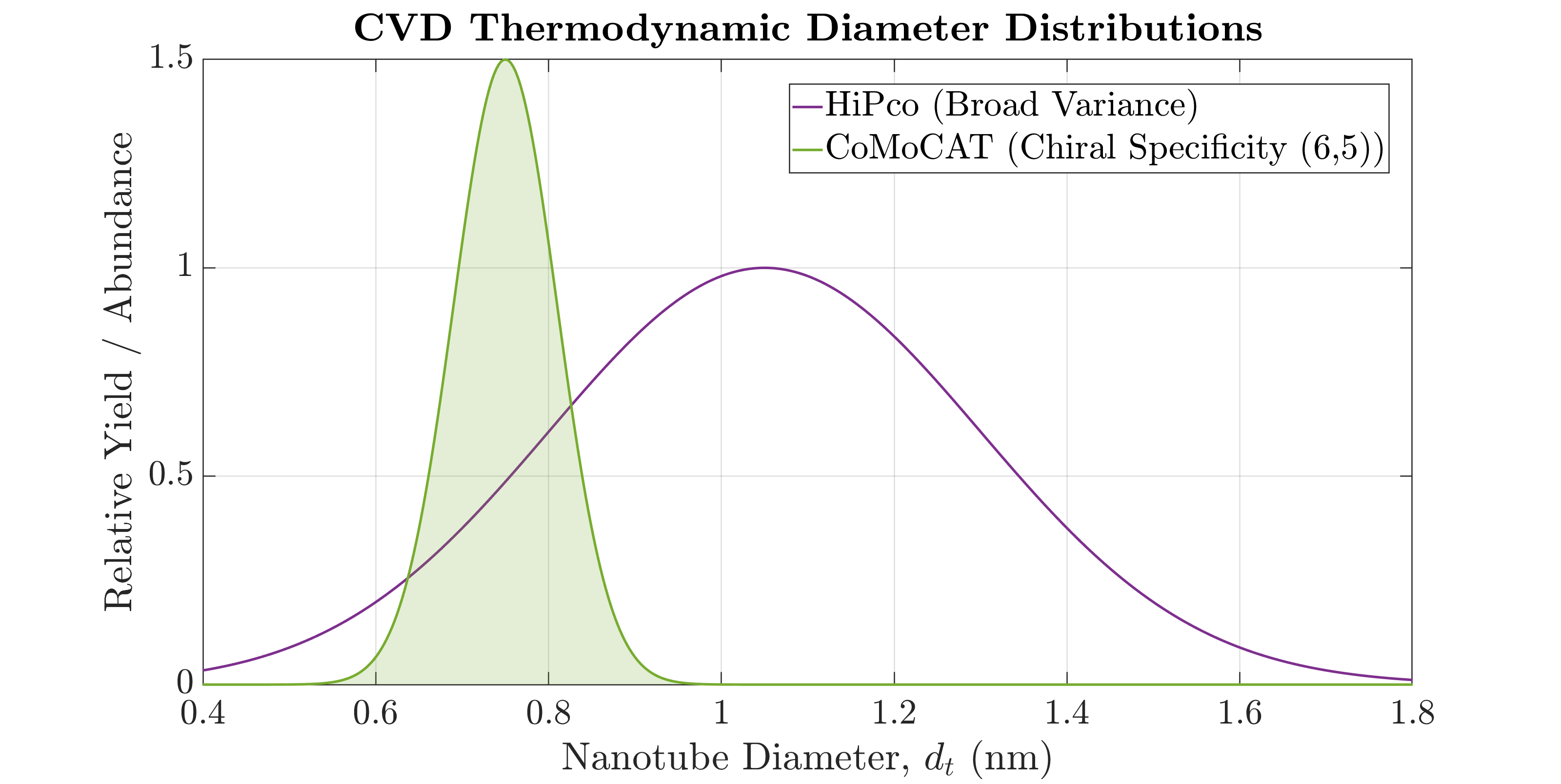}
        \caption{Schematic representation of thermodynamic diameter distributions of SWCNTs grown via HiPco (broad variance) versus CoMoCAT (highly narrow chiral specificity).}
        \label{fig:cvd_yield}
    \end{minipage}\hfill
    \begin{minipage}{\dimexpr0.5\textwidth-0.1in\relax}
        \centering
        \includegraphics[width=1\linewidth]{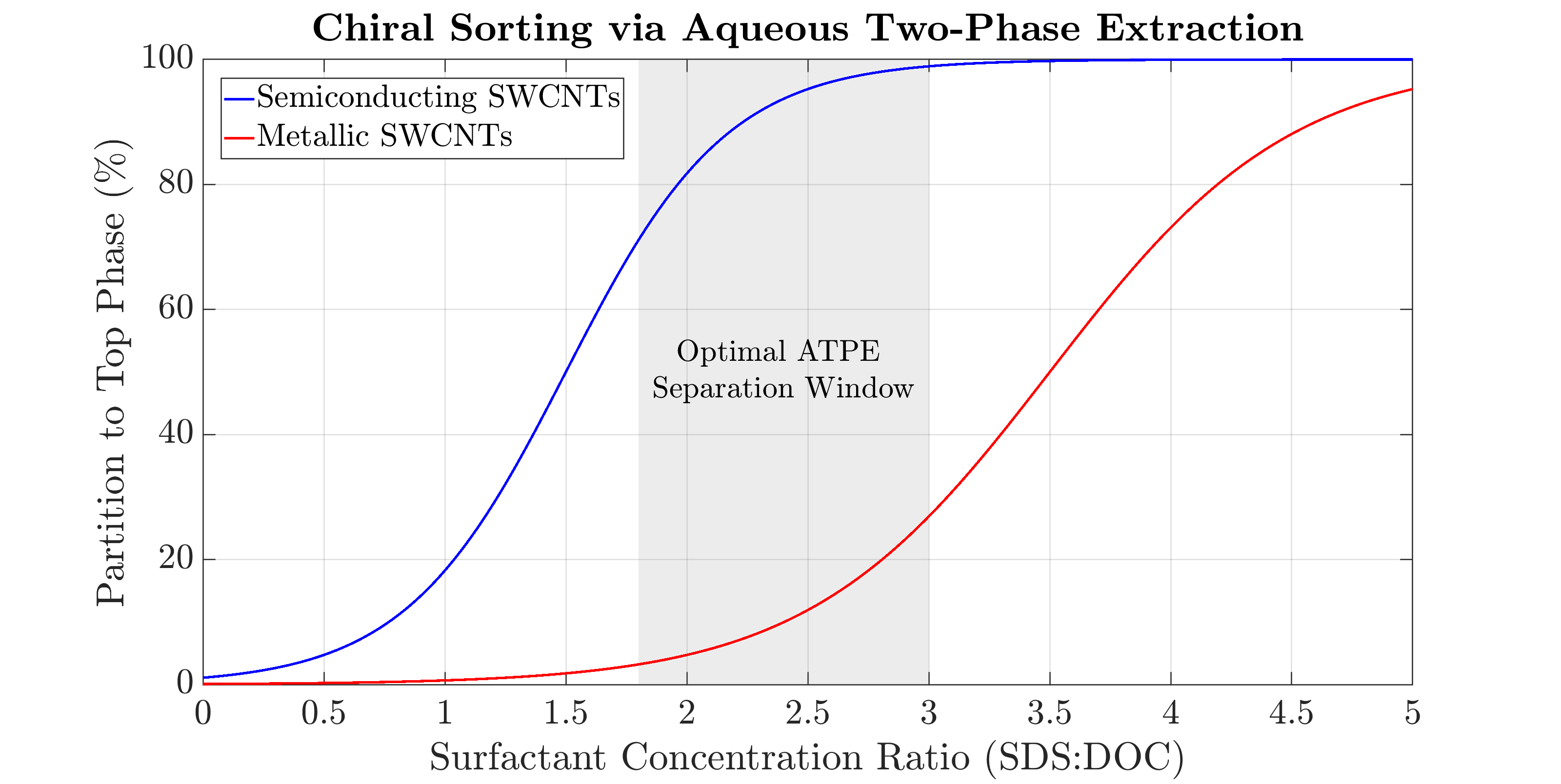}
        \caption{Schematic phase distribution and partition coefficients of metallic versus semiconducting SWCNTs in an Aqueous Two-Phase Extraction (ATPE) gradient.}
        \label{fig:atpe_partition}
    \end{minipage}
\end{figure*}

Translating exceptional transport in isolated nanotubes to wafer-scale microprocessors requires reconciling bottom-up, localized thermodynamic synthesis with rigid, top-down deterministic placement. This section evaluates thermodynamic synthesis and integration paradigms for high-density VLSI.

\subsection{From Physical Vapor Deposition to Catalytic CVD}

While initial SWCNT synthesis relied on physical vapor deposition (arc discharge, laser ablation) \cite{guo1995catalytic}, commercial VLSI integration requires Catalytic Chemical Vapor Deposition (CVD) \cite{dai2001carbon}, nucleating tubular graphene lattices from hydrocarbon precursors over transition-metal nanoparticles at $500$--$1200^\circ$C. Because SWCNT diameter is set by catalyst particle size, bimetallic templates give indirect structural control. The CoMoCAT process, for instance, uses a Cobalt-Molybdenum catalyst to thermodynamically restrict particle aggregation \cite{alvarez2001synergism}, preferentially enriching the $(6,5)$ chirality to $\sim$40--50\% of the as-grown population, as originally established in Bachilo \emph{et~al.}'s 2003 \emph{JACS} report (illustrated qualitatively via the schematic in Fig.~\ref{fig:cvd_yield})~\cite{bachilo2003controlled}. High-Pressure Carbon Monoxide (HiPco) synthesis offers higher volumetric yield \cite{nikolaev1999gas}, but its broad diameter variance makes CoMoCAT preferable for high-purity logic extraction.

\subsection{Purification, DGU, and Aqueous Two-Phase Extraction (ATPE)}

As-grown batches invariably contain structural heterogeneities, residual catalyst, and a statistical mix of metallic and semiconducting tubes that standard chemical oxidation cannot separate, demanding advanced fluidic sorting. The industry has transitioned from batch-limited Density-Gradient Ultracentrifugation (DGU)~\cite{arnold2006sorting} to scalable Aqueous Two-Phase Extraction (ATPE). Because metallic and semiconducting SWCNTs have distinct hydration enthalpies and polarizabilities, they partition into separate polymer phases, with semiconducting tubes migrating into the PEG phase and metallic tubes into the Dextran phase, as illustrated qualitatively in the schematic of Fig.~\ref{fig:atpe_partition}, with empirical implementations routinely achieving $>99.99\%$ semiconducting purity \cite{li2025precise,sims2024single}. Structural integrity and chiral fraction are then verified via Raman spectroscopy, using the $I_D/I_G$ ratio for defects and the radial breathing mode ($\omega_{RBM}\propto1/d_t$) for chiral dominance \cite{dresselhaus2005raman}.

A practically important consequence of probabilistic metallic-tube statistics is that required purity scales with array size $N$. The probability of at least one metallic tube surviving in the array is $P_{\mathrm{fail}}=1-p^N$. For a 100-tube/$\mu$m array in a 2~$\mu$m standard cell ($N=200$) at 99.99\% purity, $P_{\mathrm{fail}}\approx2\%$ per cell. Raising $p$ to six-nines (0.999999) reduces this to $\approx0.020\%$, but compounding across a billion-transistor die renders strict hardware-level purity economically unfeasible for massive arrays. Modern VLSI integration instead relies on resilient circuit architecture. Layout methodologies that pair logic gates so they can tolerate metallic shorts, building on earlier resilient-circuit-design work~\cite{patil2008design}, culminated in the DREAM approach (Designing Resiliency Against Metallic CNTs) introduced by Hills~\emph{et~al.}, who demonstrated fully functional microprocessors from commercial 99.99\% (four-nines) CNT material using this technique. This shows that circuit design is as critical as fluidic purification for CNTFET viability~\cite{hills2019modern}.

\subsection{Deterministic Placement via Dielectrophoresis (DEP)}

Once synthesized and purified, nanotubes must be deterministically aligned across source/drain electrodes via alternating-current (AC) dielectrophoresis (DEP). The time-averaged DEP force on a 1D cylindrical nanotube is:

\begin{equation}
\langle F_{\mathrm{DEP}} \rangle = \frac{\pi d_t^2 L}{4}\, \varepsilon_m\, \mathrm{Re}[K(\omega)]\, \nabla |E_{\mathrm{rms}}|^2,
\label{eq:DEP}
\end{equation}

where $\varepsilon_m$ is the medium permittivity and $K(\omega)=(\varepsilon_p^*-\varepsilon_m^*)/\varepsilon_m^*$ is the Clausius--Mossotti factor in the needle limit~\cite{Jones1995,krupke2003separation}. Conventional DEP struggles to reach the ultra-high densities ($>100$~CNTs/$\mu$m) needed to outperform silicon FinFETs and is prone to bundling at higher ink concentrations ($>1$~mg/L) \cite{krupke2003separation}. State-of-the-art foundries instead use Dimension-Limited Self-Alignment (DLSA) with optimized rheological evaporation dynamics on functionalized substrates, achieving aligned arrays exceeding $120$~tubes/$\mu$m~\cite{liu2020aligned}.

\subsection{Top-Down Patterning: Lithography and Screen Printing}

Standard photoresist processing (spin-coat, bake, EUV/deep-UV expose, develop) is governed by the Rayleigh criterion ($CD=k_1\lambda_{\mathrm{lith}}/NA$). Srimani~\emph{et~al.} showed that unmodified 193~nm immersion lithography patterns CNTFET arrays down to $CD\approx40$~nm ($k_1=0.28$, $NA=1.35$)~\cite{srimani2019immersion}, which is sufficient for first-generation gate and contact patterning. Sub-10~nm production features would instead require EUV ($\lambda_{\mathrm{lith}}=13.5$~nm) lithography with multiple patterning, along with developer-compatibility verification against the SWCNT lattice.

For macro-electronic applications such as flexible displays, additive screen printing of a viscous SWCNT ink offers a lower-cost alternative, but strong van der Waals attraction (Hamaker constant $A\approx4\times10^{-20}$~J~\cite{girifalco2000carbon}) drives bundling and agglomeration-disrupted percolation, limiting this route to macro-electronic rather than logic-grade applications~\cite{sun2013review}.

\subsection{Monolithic 3D Stack Engineering}

The low thermal budget of SWCNT processing also opens a scaling axis unavailable to lithography or screen printing alone: vertical device stacking. As 2D planar scaling exhausts its limits, the industry is pivoting toward monolithic three-dimensional (3D) integration \cite{batude20113d}. Because SWCNTs can be deposited, aligned, and metallized entirely below $400^\circ\text{C}$, they inherently support Back-End-Of-Line (BEOL) processing, enabling monolithic logic layers directly atop pre-existing silicon without thermal degradation \cite{shulaker2017three, yuan2023three}. Dense CNT arrays also ease the thermal runaway typical of 3D ICs, since their axial thermal conductivity surpasses conventional thermal interface materials, reaching $\sim$1000--6000~W/mK for individual SWCNTs~\cite{pop2006thermal} and $\sim$50--200~W/mK for aligned CNT networks~\cite{marconnet2011thermal}. 

Before taking advantage of these 3D integration benefits~\cite{wei2001reliability}, the underlying CNTFETs must be successfully engineered with low-resistance contacts, scaled dielectrics, and reliable doping schemes. We address these device-level integration challenges in Section~\ref{sec:integration}.

\section{Device Integration and Circuit Architecture}
\label{sec:integration}

With synthesis, purification, and wafer-scale placement resolved (Section~\ref{sec:manufacturing}), the remaining question is how placed tubes become working circuits, which requires assembling them into functional, complementary logic gates, or CMOS. Traditional silicon CMOS technology relies on the highly symmetric, paired operation of $n$-type and $p$-type MOSFETs to ensure that steady-state static power consumption is virtually zero. Replicating this complementary architecture using carbon nanotubes is a genuinely difficult physical challenge, requiring extreme precision in metal-work-function engineering, interfacial dielectric scaling, and chemical doping.

\subsection{Contact Resistance and the Quantum Transmission Limit}

In bulk silicon processing, highly doped source and drain extension regions ($>10^{20} \text{ cm}^{-3}$)~\cite{taur2009fundamentals} are ion-implanted to narrow the Schottky barrier at the metal-semiconductor interface, allowing for efficient field-emission tunneling. Because the atomic thickness of a SWCNT prevents direct substitutional ion implantation without catastrophically destroying the $sp^2$ transport channel, the metal contacts must be deposited directly onto the intrinsic carbon lattice.

Consequently, the contact resistance ($R_c$) is dominated by the Schottky barrier height ($\Phi_B$). For a 1D ballistic channel with two degenerate conducting modes, the fundamental quantum limit of contact resistance follows the Landauer formula ($R_{Q} = h/(4e^2 T) \approx 6.45~\text{k}\Omega/T$), where $T$ is the transmission probability. To maximize transmission ($T \to 1$), the work function of the contact metal ($\Phi_M$) must perfectly align with the energy bands of the SWCNT.

\begin{figure*}[!t]
    \centering
    \begin{minipage}{\dimexpr0.5\textwidth-0.1in\relax}
        \centering
        \includegraphics[width=1\linewidth]{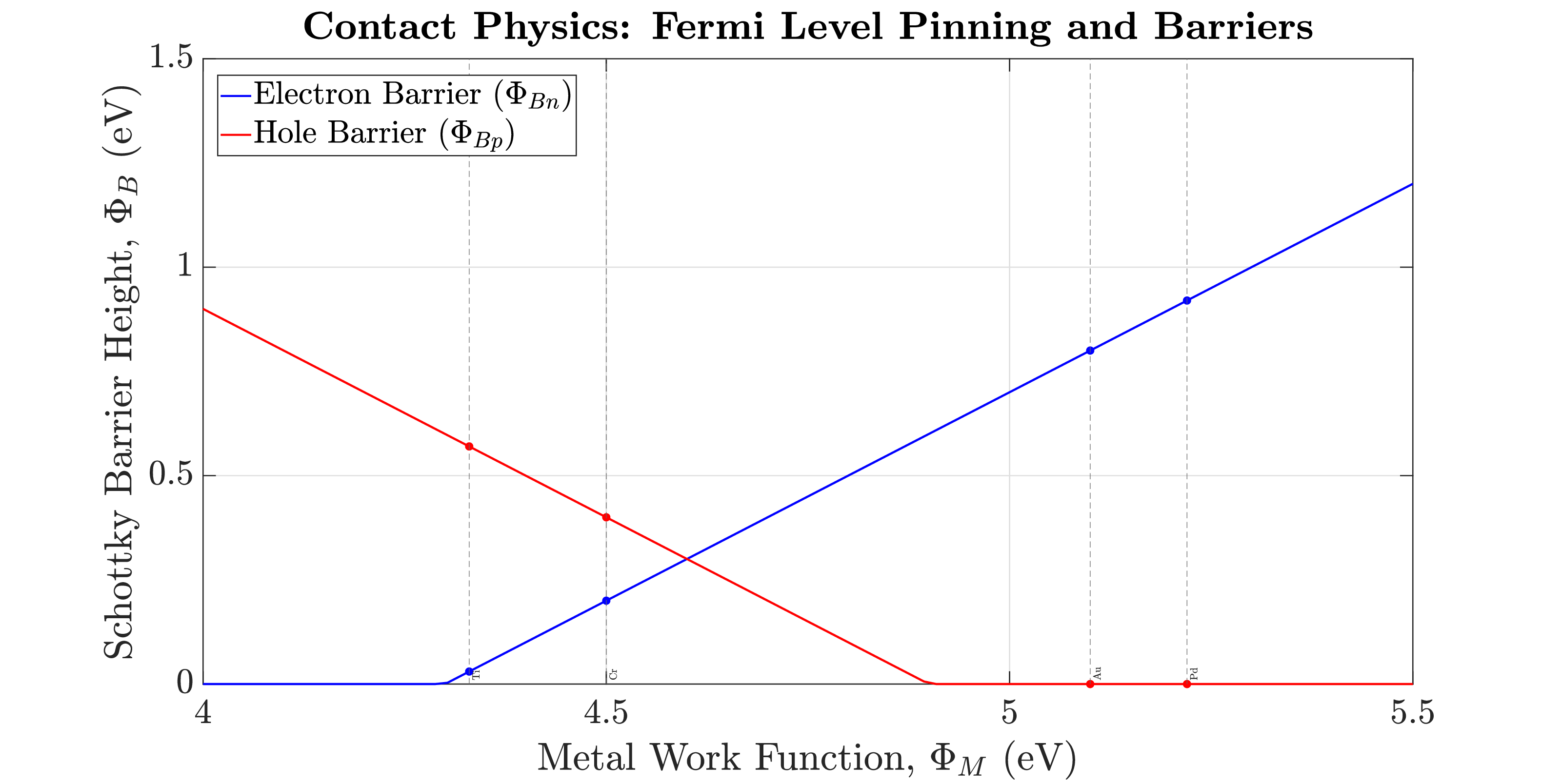}
        \caption{Schematic representation of Schottky barrier heights ($\Phi_B$) as a function of contact metal work function ($\Phi_M$), illustrating optimal Fermi level alignment for transparent hole injection utilizing Palladium.}
        \label{fig:schottky}
    \end{minipage}\hfill
    \begin{minipage}{\dimexpr0.5\textwidth-0.1in\relax}
        \centering
        \includegraphics[width=1\linewidth]{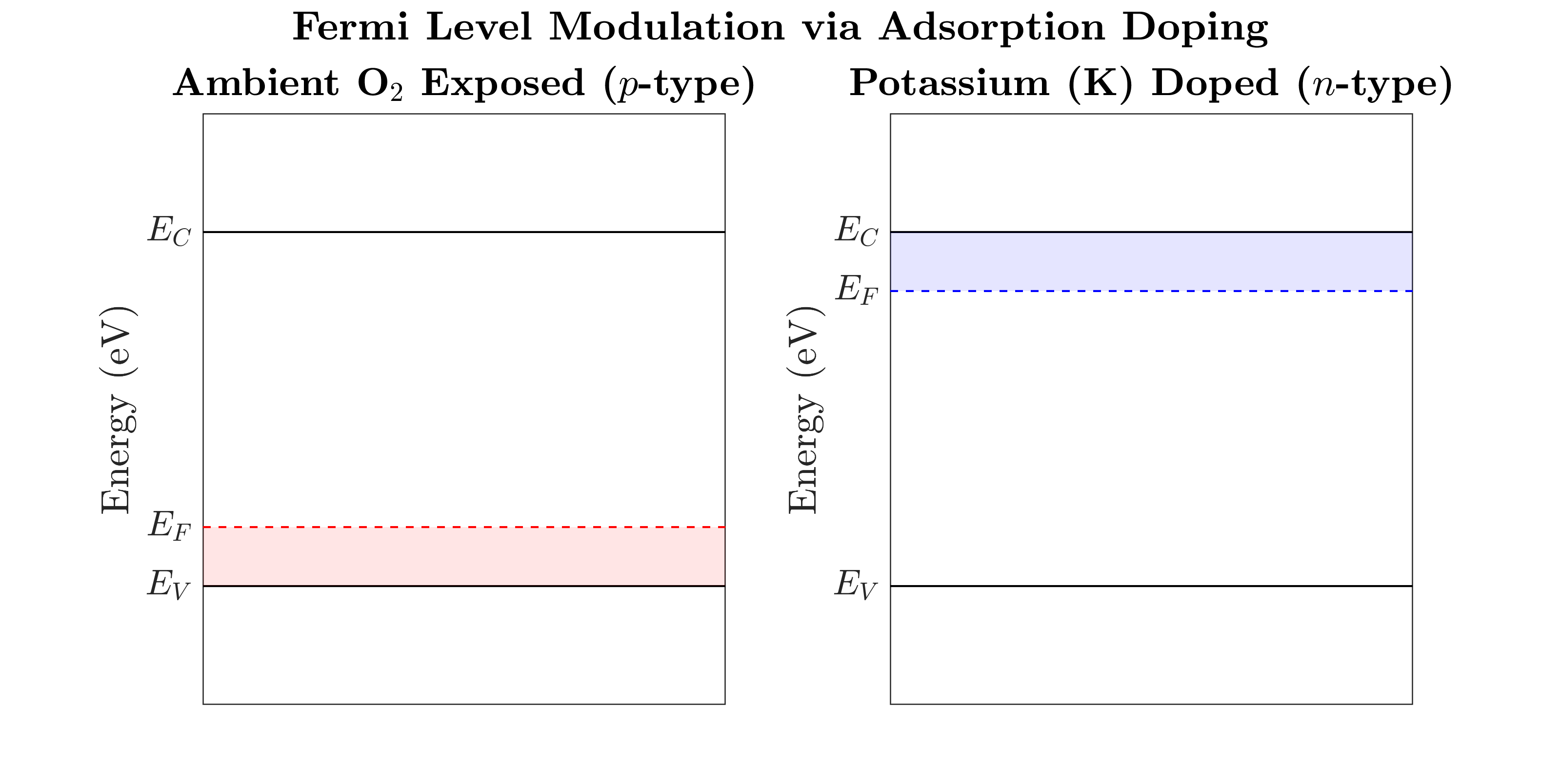}
        \caption{Energy-band schematic showing the shift from ambient $p$-type behavior to $n$-type operation via potassium adsorption doping.}
        \label{fig:band_diagrams}
    \end{minipage}
\end{figure*}

For hole injection ($p$-type operation), the barrier height is $\Phi_{Bp} = IP - \Phi_M$, where $IP \approx 5.05~\text{eV}$ is the ionization potential for the representative $(13,0)$ nanotube (consistent with the quasiparticle band-edge alignment reported in~\cite{jiang2007quasiparticle}). As demonstrated in Fig.~\ref{fig:schottky}, utilizing high-work-function metals such as Palladium (Pd, $\Phi_M \approx 5.1\text{--}5.2 \text{ eV}$) effectively yields a negative or near-zero barrier, resulting in highly transparent, ballistic $p$-FET contacts \cite{javey2003ballistic}. 

Cao~\emph{et~al.} demonstrated end-bonded metal--carbon contacts, in which the metal is covalently bonded to the open tube end rather than the sidewall. Validating the fundamental limit, they reported an un-normalized contact resistance of $36~\text{k}\Omega$ and delivered drive currents approaching $\sim$15~$\mu$A per tube \cite{cao2015end}. Though still above the quantum limit, this metallurgical junction gives length-independent $R_c$, enabling functional scaling to 5~nm gate lengths without severe current degradation.

Conversely, for electron injection ($n$-type operation), the barrier is $\Phi_{Bn} = \Phi_M - \chi$, where $\chi \approx 4.3~\text{eV}$ is the electron affinity of the SWCNT~\cite{jiang2007quasiparticle,javey2003ballistic}. Low-work-function metals like Ti or Al historically oxidized and degraded the interface \cite{javey2003ballistic}, but utilizing Scandium ($\Phi_M \approx 3.3$~eV) as the direct contact metal, as explicitly demonstrated in the definitive doping-free $n$-FET architectures of Zhang~\emph{et~al.}~\cite{zhang2007doping}, alongside Yttrium contacts under ultra-high vacuum \cite{ding2009yttrium}, yields near-ballistic $n$-FETs. This established the symmetric work-function engineering foundation for CMOS logic.

Achieving $T\to1$ also depends on the electrostatic sharpness of the junction, governed by Fowler-Nordheim triangular-barrier tunneling through the finite Schottky barrier $\Phi_B$. The local junction field is approximated here via the same natural scaling length that governs channel electrostatics ($F_c \approx \Phi_B/(e\,\lambda_{\mathrm{GAA}})$), serving strictly as a zeroth-order geometrical approximation. As established by L\'{e}onard and Tersoff, a rigorous treatment requires the junction-specific electrostatics of a 1D nanotube contact. These differ fundamentally from bulk 3D Schottky theory because the depletion width scales exponentially with inverse doping, and the charge decays logarithmically over a length defined by the tube radius~\cite{leonard1999novel}. It should be noted that stacking the idealized channel scaling approximation on top of the contact field approximation omits critical multi-dimensional fringing effects. Still, substituting this field approximation into WKB tunneling yields an exponential relationship for contact resistance:

\begin{equation}
R_{c} \approx \frac{h}{4e^2} \exp\!\left[\frac{4\lambda_{\mathrm{GAA}}\sqrt{2m^*_b}\,\Phi_{B}^{1/2}}{3\hbar}\right],
\label{eq:rc_schottky}
\end{equation}

where $m^*_b$ is the effective carrier tunneling mass at the metal--SWCNT Schottky interface (distinct from the SWCNT band effective mass and the silicon mass used in Eq.~\eqref{eq:wkb}). This makes it explicit why minimizing $\Phi_B$ through work-function alignment and minimizing $\lambda_{\mathrm{GAA}}$ through increased gate coupling both reduce contact resistance exponentially, connecting the Schottky-barrier engineering discussed here with the electrostatic scaling discussed in Section~\ref{sec:physics}~\cite{leonard1999novel, javey2003ballistic}.

Physical contact geometry and length scaling matter independently of barrier height \cite{franklin2010length}. Standard "side-bonded" top contacts, for example, suffer a van der Waals tunneling gap that bottlenecks transmission below $L_c\approx10$~nm \cite{franklin2010length}. Cao~\emph{et~al.}'s end-bonded contacts remove this gap entirely, while Qiu~\emph{et~al.}'s graphene-contacted 5~nm devices (Section~\ref{sec:transport}) independently validate length scaling via a different route \cite{qiu2017scaling}. Dirac-source architectures have further been shown to steepen subthreshold swing beyond the thermionic limit \cite{qiu2018dirac}. These represent two independently validated, length-scalable strategies for maintaining low contact resistance into the single-digit-nanometer regime.

\subsection{Dielectric Interface and High-$\kappa$ Nucleation}

Low-resistance contacts address only half the electrostatic picture. The other half is the gate dielectric itself. Maximizing electrostatic coupling over the 1D channel without severe Fowler-Nordheim gate leakage requires ultra-thin high-$\kappa$ dielectrics (HfO$_2$, Al$_2$O$_3$) deposited via Atomic Layer Deposition (ALD), whose efficiency is measured by the Equivalent Oxide Thickness of Eq.~\eqref{eq:EOT_def}. However, the pristine, fully-bonded SWCNT lattice lacks the dangling bonds ALD precursors (e.g., trimethylaluminum) need to nucleate, so direct ALD produces non-uniform, "beaded" dielectric islands that fail to pinch off the channel.

To resolve this, the field has developed non-covalent functionalization techniques. By exposing the SWCNT to a highly controlled NO$_2$ gas environment prior to ALD, a uniform monolayer of nucleation sites is non-covalently adsorbed onto the carbon lattice without breaking the underlying $sp^2$ transport bonds~\cite{farmer2006atomic,lu2006atomic}. This permits the conformal growth of sub-3 nm high-$\kappa$ dielectrics, driving the EOT below 1 nm and virtually eliminating short-channel effects.

\subsection{Intrinsic Ambipolarity and Reversible Chemical Doping}

With contacts and dielectric addressed, the last device-level ingredient for complementary logic is control over carrier polarity. A perfectly passivated, intrinsic SWCNT utilizing mid-gap contacts exhibits ambipolar transport behavior, conducting both electrons and holes symmetrically depending on the polarity of the applied gate bias. However, in ambient laboratory environments, un-passivated CNTFETs invariably demonstrate strongly unipolar $p$-type characteristics.

Collins~\emph{et~al.} traced this $p$-type dominance to extreme oxygen sensitivity \cite{collins2000extreme}. Ambient O$_2$ and water vapor adsorb on the nanotube surface and at the contact interfaces, and their electronegativity pins $E_F$ close to the valence band.

Two doping methodologies enable deterministic $n$-type regions for complementary logic. Alkali-metal adsorption (e.g., Potassium) donates electrons, shifting $E_F$ upward to form an $n$-FET~\cite{derycke2001intramolecular} with $I_{ON}/I_{OFF}$ up to $10^5$~\cite{derycke2002controlling} (Fig.~\ref{fig:band_diagrams}), though reversibility on oxygen exposure necessitates hermetic encapsulation \cite{collins2000extreme}. Alternatively, vacuum annealing at 200--300$^\circ$C falls within the BEOL thermal budget and thermally desorbs adsorbed O$_2$/H$_2$O to yield $n$-type behavior non-destructively~\cite{derycke2002controlling}. Selectively re-exposing the substrate to oxygen gradients then restores $p$-type character locally, enabling tunable CMOS definition.

\subsection{Intramolecular Logic and Microprocessor Architecture}

Selective segment-by-segment doping of a single SWCNT enables dense intramolecular logic with intrinsic voltage gain $|A_v|>1$ \cite{derycke2001intramolecular,zhang2007doping}. Following the first working carbon-nanotube computer~\cite{shulaker2013carbon}, this line of work has culminated in the RV16X-NANO, a fully functional 16-bit RISC-V microprocessor built from over 14,000 CNTFETs \cite{hills2019modern}, which remains the strongest proof to date that CNT CMOS logic scales beyond isolated devices and ring oscillators \cite{bachtold2001logic} into general-purpose computation. This was subsequently reinforced by CNTFET fabrication within a commercial 200~mm silicon foundry \cite{bishop2020fabrication}. Converting this into a qualified commercial process, however, still requires answering whether these devices survive continuous-operation bias/thermal stress, and whether purification and assembly scale to foundry volumes, the first of which is addressed next.

\subsection{Reliability: Gate Hysteresis and Bias Temperature Instability}
\label{sec:bti}

Non-passivated devices measured in ambient air exhibit a clockwise gate-voltage hysteresis window of 0.2--0.5~V, from charge trapping/de-trapping at adsorbed hydroxyl, water, and slow dielectric trap sites \cite{lu2006atomic,collins2000extreme}. This drifts $V_T$ with sweep history and temperature, which is incompatible with static CMOS noise margins. Conformal encapsulation in a second ALD Al$_2$O$_3$ layer ($\sim$10~nm) blocks water adsorption and passivates the dominant hydroxyl traps, reducing hysteresis below 50~mV \cite{lu2006atomic}, a step that must be built into any commercial BEOL CNTFET flow.

A distinct, longer-term concern is Bias Temperature Instability (BTI), a cumulative $V_T$ shift under prolonged bias/temperature stress, qualitatively similar to NBTI/PBTI in silicon and 2D-material FETs. While initial mitigation strategies focused on interface optimization, Sun~\emph{et~al.}~\cite{sun2025bti} proposed comprehensive CNT/high-$\kappa$ stability frameworks, and a preprint by Yu~\emph{et~al.}~\cite{yu2024overcoming} reports that AC pulsed operation with nitride encapsulation extends NBTI time-to-failure in foundry-fabricated CNTFETs, a result that awaits independent peer-reviewed confirmation. However, a unified, foundry-calibrated reaction-diffusion model analogous to silicon qualification does not yet exist. The ambient $p$-type character of unpassivated CNTFETs suggests predominant NBTI susceptibility, mirroring $p$-type silicon MOSFETs, and that established silicon mitigation strategies (nitrogen passivation, ALD oxidant chemistry optimization) may be transferable. Quantifying and mitigating BTI at foundry-qualification level remains an open challenge for commercial CNTFET adoption.

\section{Techno-Economic Analysis and Commercial Foundry Feasibility}
\label{sec:economics}
 
With the device physics, materials synthesis, and basic circuit integration pathways outlined in the previous sections, the ultimate question is whether any of this is commercially actionable. The IRDS 2022 roadmap explicitly identifies "1D and 2D channel materials," including SWCNTs, as the primary candidate technologies to sustain transistor density scaling beyond the 2~nm silicon node \cite{irds2022}. This section evaluates the economic and engineering barriers governing the transition from this recognition to foundry adoption, grounding the analysis in publicly available semiconductor industry data, peer-reviewed yield modeling, and demonstrated BEOL integration results.
 
\subsection{The IRDS Roadmap and the Post-Silicon Imperative}
 
Moore's Law projected doubling transistor density every two years \cite{moore1965cramming}. This held for decades even as node cost grew super-exponentially, with a leading-edge EUV fab now costing \$15--\$20~billion~\cite{irds2022,waldrop2016chips}. The IRDS projects bulk silicon MOSFETs will reach fundamental electrostatic limits at the 1~nm EOT node ($\sim$2026--2028), beyond which its ``Beyond CMOS'' section identifies three requirements for successor technologies, namely sub-1~nm effective body thickness, high intrinsic carrier velocity, and BEOL-compatible ($<400^\circ$C) deposition. SWCNTs satisfy all three, as demonstrated throughout Sections~\ref{sec:physics}--\ref{sec:integration}.
 
\subsection{Capital Expenditure and the Defect Density Problem}
 
Despite the physical superiority of the CNTFET, commercial foundry adoption is dictated by two coupled economic constraints. The first is the up-front capital expenditure (CapEx) required to qualify a new process module, and the second is the manufacturing yield determining the cost per functional die. A complete greenfield transition to CNT-specific tools is economically prohibitive. However, because BEOL processing operates below the thermal degradation limits of the underlying silicon ($<400^\circ\text{C}$), CNT modules can be introduced as sequential add-ons to established logic flows rather than requiring a multi-billion-dollar greenfield fab construction \cite{irds2022}.

Yield is the second constraint. Extending Murphy's original yield formulation \cite{murphy1964cost}, the negative binomial model gives:

\begin{equation}
Y_{\mathrm{die}} = \left(1 + \frac{A_{\mathrm{die}} \cdot D_{\mathrm{CNT}}}{\alpha}\right)^{-\alpha},
\label{eq:defect}
\end{equation}

where $A_{\mathrm{die}}$ is the die area, $D_{\mathrm{CNT}}$ is the fatal defect density (dominated by residual metallic SWCNTs), and $\alpha$ is the spatial clustering parameter of the defect distribution (typically $\alpha = 1$--$3$ for mature CMOS processes)~\cite{Stapper1983}. $D_{\mathrm{CNT}}$ is highly sensitive to unremoved metallic SWCNTs, and the CoMoCAT/ATPE pipeline's ability to drive metallic concentration below 0.01\% mathematically suppresses it \cite{li2025precise,sims2024single}. This underscores fluidic purification as the primary economic lever governing CNTFET viability, not merely a materials challenge.
 
\subsection{Demonstrated Manufacturing Progress and the Path to Foundry Adoption}
 
The most compelling economic evidence for CNT viability is Bishop~\emph{et~al.}'s demonstration \cite{bishop2020fabrication} that standard 200~mm silicon foundry toolsets apply to CNTFET fabrication with minimal modification, bridging the university-scale RV16X-NANO baseline \cite{hills2019modern} to a manufacturable process. This is reinforced by GHz-class CNT thin-film circuits~\cite{zhong2018gigahertz}, 193~nm immersion-lithography-patterned arrays~\cite{srimani2019immersion}, and monolithic 3D CNT-logic-on-silicon/DRAM integration~\cite{yuan2023three}, together showing the steps validated individually in Sections~\ref{sec:manufacturing}--\ref{sec:integration} also function together at the wafer level foundry qualification requires.

\subsection{Environmental Health, Safety, and Operational Variability}

Beyond classical yield constraints, introducing novel carbon nanomaterials into established foundries triggers stringent Environmental Health and Safety (EHS) regulations. The respiratory toxicology of aerosolized SWCNTs, often analogized to asbestos fibers depending on tube length and agglomeration~\cite{poland2008carbon}, dictates that dry powder handling is strictly prohibited in standard cleanrooms. Wafer-scale integration mandates enclosed, liquid-phase functionalization and deposition pipelines to mitigate occupational exposure, adding infrastructural overhead to the BEOL transition. Finally, foundry qualification demands rigorous parameter matching across industrial temperature ranges ($-40^\circ\text{C}$ to $125^\circ\text{C}$). The $V_T$ variability inherent in slight remaining chirality distributions currently limits the viability of dense SWCNT arrays for highly symmetric SRAM cells.

Together, IRDS recognition of CNTs as the primary beyond-CMOS candidate, demonstrated commercial-foundry compatibility, and the BEOL integration pathway constitute a low-risk adoption model. Rather than replacing the multi-trillion-dollar silicon infrastructure, CNTFET modules are positioned as incremental performance accelerators within the BEOL stack, sustaining density scaling through the early post-silicon era.

\section{Conclusion}
\label{sec:conclusion}

This review has developed a consistent argument across Sections~\ref{sec:physics}--\ref{sec:economics}. Specifically, 1D confinement collapses the electron-phonon scattering phase space to yield near-ballistic transport alongside a favorable scaling length and quantum-capacitance-limited switching energy. We extended these bounds to derive an exact mathematical equivalence point between 1D nanotubes and 2D transition metal dichalcogenides. This proves that array densities would theoretically need to exceed 1720 tubes per micrometer to match the capacitance of a 2D sheet, a density that exceeds even the physical close-packing limit ($\approx 1000$ tubes/µm), confirming this gap cannot be closed by density alone. This highlights why superior carrier velocity---rather than array density---must carry the CNTFET's competitive advantage, consistent with the caveat in Section~\ref{sec:transport} that idealized coaxial {GAA} electrostatics remain a theoretical target rather than a demonstrated property of any device discussed in this review. Furthermore, comparing our theoretical framework against the 5~nm experimental data of Qiu~\emph{et~al.}~\cite{qiu2017scaling} strongly supports the hypothesis that direct source-to-drain tunneling becomes the binding constraint at sub-5~nm dimensions rather than classical electrostatics. On the manufacturing side, the CoMoCAT/ATPE pipeline delivers the purity that defect-tolerant layouts such as DREAM require, DEP and DLSA provide deterministic wafer-scale placement, and Schottky-engineered Pd/Sc contacts together with non-covalently functionalized high-$\kappa$ dielectrics complete a working, BEOL-compatible complementary logic technology, as the RV16X-NANO microprocessor and its subsequent commercial-foundry fabrication demonstrate.

Several open challenges still remain. The electrostatic and power-efficiency bounds discussed in Section~\ref{sec:transport} represent idealized, single-tube asymptotes. Real interconnect and fan-out parasitics will heavily shape the final standard-cell metrics, although the fundamental per-tube quantum-capacitance advantage still holds. The $>$99.99\% purity from CoMoCAT/ATPE (Section~\ref{sec:manufacturing}) is chip-scale viable only with a defect-tolerant layout such as DREAM~\cite{hills2019modern}, and chirality polydispersity still limits bandgap uniformity for tight-tolerance analog/RF use. Gate hysteresis is largely solved by Al$_2$O$_3$ encapsulation~\cite{lu2006atomic}, but bias temperature instability remains at an early characterization stage~\cite{sun2025bti}, well short of silicon's foundry-calibrated reliability models. Finally, the techno-economic case rests on device-level BEOL demonstrations~\cite{bishop2020fabrication,yuan2023three} rather than decades of wafer-scale qualification data, and the yield model of Eq.~\eqref{eq:defect} does not yet capture correlated defects such as incomplete DEP alignment.

Rather than treating the device physics as entirely solved, these open challenges mark the critical intersections of fundamental physics, reliability modeling, and integration milestones needed to qualify a commercial CNTFET process. By collapsing the electron-phonon scattering phase space, SWCNTs offer a physically rigorous path to sustain density scaling deep into the sub-10~nm regime, and with low-temperature 3D BEOL integration already demonstrated, the carbon nanotube has moved decisively from a theoretical curiosity to a credible, architecturally scalable successor to silicon.

\section*{Acknowledgment}

The authors gratefully acknowledge Sradha Mishra, Sunit Sahoo and Karthik S. for their fruitful discussions regarding the underlying physics and early concepts that inspired this review.

\bibliographystyle{IEEEtran}
\bibliography{ref}

\end{document}